\documentclass[a4paper,12pt]{article}
\usepackage{amssymb,psfig}
\renewcommand{\[}{\begin{equation}}
\renewcommand{\]}{\end{equation}}
\newcommand{\intr}{\int d{\bf r} \;}

\def\epl#1#2#3{Europhys. Lett. {\bf #1}, #2 (19#3)}

\def\prb#1#2#3{Phys. Rev. B {\bf #1}, #2 (19#3)}
\def\prl#1#2#3{Phys. Rev. Lett. {\bf #1}, #2 (19#3)}

\def\equ#1{Eq.~(\ref{#1})}
\def\eqs#1#2{Eqs.~(\ref{#1}) and (\ref{#2})}
\begin{document}

\setlength{\unitlength}{1cm}
\noindent \begin{picture}(0,0)
\put(0,2.5){\noindent \sf Presented at the Sanibel Symposium 1999,}
\put(0,2){\noindent \sf February 27 -- March 5,
St. Augustine, Florida, USA.}
\put(0,1.5){\noindent \sf To appear on: International Journal of Quantum
Chemistry}
\end{picture}
\begin{center}
{\Large\bf Macroscopic Polarization \\ 
from Electronic Wavefunctions} \\

\vspace{0.5cm}
{Raffaele Resta} \\

\vspace{0.6cm}
{\it INFM -- Dipartimento di Fisica Teorica, \\
Universit\`a di Trieste, Strada Costiera 11, 
I-34014 Trieste, Italy}\\
\end{center}

\begin{abstract} The dipole moment of any finite and neutral system, having a
square-integrable wavefunction, is a well defined quantity. The same quantity
is ill--defined for an extended system, whose wavefunction invariably obeys
periodic (Born-von K\'arm\'an) boundary conditions.  Despite this fact,
macroscopic polarization is a theoretically accessible quantity, for either
uncorrelated or correlated many--electron systems: in both cases, polarization
is a rather ``exotic'' observable. For an uncorrelated---either Hartree--Fock
or Kohn--Sham---crystalline solid, polarization has been expressed and
computed as a Berry phase of the Bloch orbitals (since 1993).  The case of a
correlated and/or disordered system received a definitive solution only very
recently (1998): this latest development allows us to present here the whole
theory from a novel, and very general, viewpoint.  The modern theory of
polarization is even relevant to the foundations of density functional theory
in extended systems.  \end{abstract}

\vspace{0.2cm}

\section{Introduction}

The dipole moment of any {\it finite} $N$--electron system in its ground state
is a simple and well defined quantity. Given the many--body wavefunction
$\Psi$ and the corresponding single--particle density $n({\bf r})$ the
electronic contribution to the dipole is: \[ \langle {\bf R}
\rangle = \intr {\bf r} \, n({\bf r}) = \langle \Psi | \hat{{\bf R}} | \Psi
\rangle , \label{dipole} \] where $\hat{{\bf R}} = \sum_{i=1}^N {\bf r}_i$
(atomic Hartree units are adopted throughout). This looks very trivial, but
we are exploiting here an essential fact: the ground wavefunction of any
finite $N$--electron system is square--integrable and vanishes exponentially
at infinity; the density vanishes exponentially as well.

Considering now a macroscopic solid, the related quantity is macroscopic
polarization, which is a very essential concept in any phenomenological
description of dielectric media~\cite{Landau}: this quantity is ideally
defined as the dipole of a macroscopic sample, divided by its volume.  The
point is that, when using \equ{dipole}, the integral is dominated by what
happens at the surface of the sample: knowledge of the electronic distribution
in the bulk region is not enough to unambiguosly determine the dipole. This
looks like a paradox, since in the thermodynamic limit macroscopic
polarization must be an intensive quantity, insensitive to surface effects. 

Macroscopic polarization in the bulk region of the solid must be determined by
what ``happens'' in the bulk as well. This is the case if one assumes a model
of discrete and well separated dipoles, \`a la Clausius-Mossotti: but real
dielectrics are very much different from such an extreme model. The
valence electronic distribution is continuous, and often very delocalized
(particularly in covalent dielectrics). Most textbooks attempt at explaining
the polarization of a periodic crystal via the dipole moment of a unit cell,
or something of the kind \cite{Kittel,Ashcroft}. These definitions are
incorrect \cite{Martin74}: according to the modern viewpoint, bulk macroscopic
polarization is a physical observable {\it completely independent} from the
periodic charge distribution of the polarized crystalline dielectric.

In condensed matter physics the standard way for getting rid of undesired
surface effects is to adopt periodic Born-von K\'arm\'an
boundary conditions (BvK). Indeed, the BvK choice is mandatory in order to
introduce even the most elementary topics, such as the free--electron gas and
its Fermi energy, or the Bloch theorem \cite{Kittel,Ashcroft}. Unfortunately,
the adoption of BvK does not solve the polarization problem. In fact the
dipole {\it cannot} be evaluated as in \equ{dipole} when the wavefunction
obeys BvK: the integrals are ill--defined due to the unbounded nature of the
quantum--mechanical position operator.

For this reason macroscopic polarization remained a major challenge in
electronic structure theory for many years. The breakthrough came in 1992,
when polarization was defined in terms of the wavefunctions, not of the
charge. This definition has an unambiguous thermodynamic limit, such that BvK
and Bloch states can be used with no harm~\cite{rap73}. In the following
months a modern theory of macroscopic polarization in crystalline dielectrics
has been completely established \cite{rap78,rap_a12}, thanks to a major
advance due to R.D.  King-Smith and D. Vanderbilt \cite{King93}, who expressed
polarization in terms of a Berry phase \cite{Berry84,Berry,rap_a17}. A
comprehensive account of the modern theory exists \cite{rap_a12}. Other less
technical presentations are available as well \cite{rap82,rap_a16,Martin97a};
for an oversimplified nontechnical outline see Ref.~\cite{rap_a18}.
First--principle calculations based on this theory have been performed for
several crystalline materials, either within DFT (using various basis sets) or
within HF (using LCAO basis sets)~\cite{William}.

All of the above quoted work refers to a crystalline system within an
independent--electron formulation: the single--particle orbitals have then the
Bloch form, and the macroscopic polarization is evaluated as a Berry phase of
them. The related, but substantially different, problem of macroscopic
polarization in a correlated many--electron system was first solved by
Ort\'{\i}z and Martin in 1994~\cite{Ortiz94}. However, according to them,
polarization is defined---and computed~\cite{rap87,Ortiz95}---by means of a
peculiar ``ensemble average'', integrating over a set of different electronic
ground states: this was much later (1998) shown to be unnecessary. In
Ref.~\cite{rap100}, in fact, a simpler viewpoint is taken: the polarization of
a correlated solid is defined by means of a ``pure state'' expectation value,
although a rather exotic kind of one. By the same token, it was also possible
to define~\cite{rap_a17,rap100,rap101}---and to
compute~\cite{simulations}---macroscopic polarization in noncrystalline
systems.  

In the present work we take advantage of the most recent developments for
reconsidering the whole theory under a new light. At variance with previous
presentations we are not going to introduce {\it explicitly} the Berry phase
concept: in fact, within the formulation of Ref.~\cite{rap100}, the Berry
phase appears very much ``in disguise''. Instead, a major role is played by a
precursor work~\cite{Selloni87, brodo}, apparently unrelated either to the
polarization problem or to a Berry phase, which will be reexamined here and
used to introduce the polarization theory.

Finally, let me just mention a latest development, not to be discussed in the
present work, where ideas spawned from the polarization theory---and more
specifically from Ref.~\cite{rap100}---are used to investigate wavefunction
localization~\cite{rap107}.

\section{The ``electron in broth''}

Adopting a given choice for the boundary conditions is tantamount to defining
the Hilbert space where our solutions of Schr\"odinger's equation live.  For
the sake of simplicity, I am presenting the basic concept by means of the
one--dimensional case. For a single--particle wavefunction BvK reads
$\psi(x+L) = \psi(x)$, where $L$ is the imposed periodicity, chosen to be
large with respect to atomic dimensions.  Notice that lattice periodicity is
{\it not} assumed, and BvK applies to disordered systems as well. 

By definition, an operator maps any vector of the given Hilbert space into
another vector belonging to the same space: the multiplicative position
operator $x$ is therefore {\it not} a legitimate operator when BvK are adopted
for the state vectors, since $x \,\psi(x)$ is not a periodic function whenever
$\psi(x)$ is such. It is then obvious why \equ{dipole} cannot be used in
condensed matter theory. Of course, any periodic function of $x$ is a
legitimate multiplicative operator in the Hilbert space: this is the case {\it
e.g.} of the nuclear potential acting on the electrons.  

Before switching to the polarization problem, it is expedient to discuss an
important precursor work, apparently unrelated to the polarization problem,
where nonetheless the expectation value of the position operator plays the key
role.

Some years ago, A. Selloni {\it et al.} \cite{Selloni87} addressed the
properties of electrons dissolved in molten salts at high dilution, in a
paper which at the time was commonly nicknamed the ``electron in broth''. 
The physical problem was studied by means of a mixed quantum--classical
simulation, where a lone electron was adiabatically moving in a molten salt
(the ``broth'') at finite temperature. The simulation cell contained 32
cations, 31 anions, and a single electron. KCl was the original case study,
which therefore addressed the liquid state analogue of an F center; other
systems were studied afterwards \cite{brodo}. The motion of the ions was
assumed as completely classical, and the Newton equations of motion were
integrated by means of standard molecular dynamics (MD) techniques, though
the ionic motion was coupled to the quantum degree of freedom of the
electron. The electronic ground wavefunction was determined solving the
time--dependent Schr\"odinger's equation at each MD time step.  As usual in MD
simulations, periodic boundary conditions were adopted for the classical ionic
motion. Ideally, the ionic motion occurs in a simulation cell which is
surrounded by periodic replicas: inter-cell interactions are accounted for,
thus avoiding surface effects.  Analogously, the electronic wavefunction is
chosen in the work of Selloni {\it et al.} to obey BvK over the simulation
cell, and therefore features periodic replicas as well. A plot of such an
electronic distribution, in a schematic one--dimensional analogue, is given
in Fig.~1. 

\begin{figure}[t] \centerline{\psfig{file=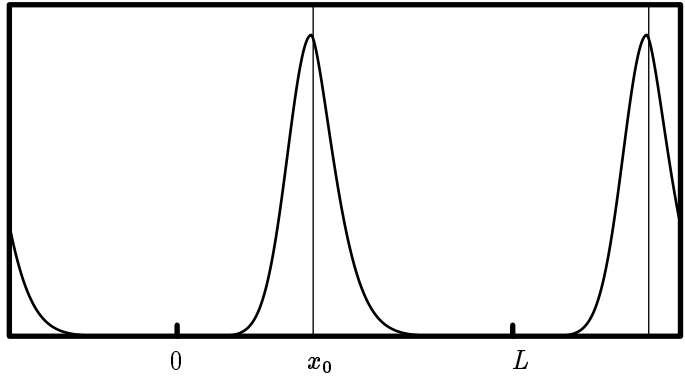,width=9cm}}
\caption[1]{The distribution $|\psi(x)|^2$ of a single--particle orbital
within periodic Born--von--K\`arm\`an boundary conditions}
\label{loc} \end{figure}

One of the main properties investigated in Ref.~\cite{Selloni87} was the
electronic diffusion, where the thermal ionic motion is the driving agent
(within the adiabatic approximation). In order to perform this study, one has
to identify first of all where the ``center'' of the electronic distribution
is.  Intuitively, the distribution in Fig.~1 appears to have a ``center'',
which however is defined only modulo the replica periodicity, and furthermore
{\it cannot} be evaluated  simply as in \equ{dipole}, {\it i.e.} $\langle x
\rangle = \int d x \, x |\psi(x)|^2$, precisely because of BvK. Selloni {\it
et al.} solved the problem by means of a very elegant and far--reaching
formula, presented below. The work of Ref.~\cite{rap100} can be regarded as
the many--body generalization of it.

\section{The main formula: One electron}

According to Refs.~\cite{rap100,Selloni87,brodo,rap107}, the key quantity for
dealing with the position operator within BvK is the dimensionless complex
number ${\mathfrak z}$, defined as: \[ {\mathfrak z} = \langle \psi  | {\rm
e}^{i\frac{2\pi}{L}x} | \psi \rangle = \int_0^L \! dx \; {\rm
e}^{i\frac{2\pi}{L}x} |\psi(x)|^2 , \label{z} \] whose modulus is no larger
than 1. The most general electron density, such as the one depicted in Fig. 
1, can always be written as a superposition of a function $n_{\rm loc}(x)$,
normalized over $(-\infty, \infty)$, and of its periodic replicas:
\begin{equation} |\psi(x)|^2 = \sum_{m = -\infty}^\infty n_{\rm loc} (x - x_0
-mL) . \label{replicas} \end{equation} Both $x_0$ and $n_{\rm loc} (x)$ have a
large arbitrariness: we restrict it a little bit by imposing that $x_0$ is the
center of the distribution, in the sense that $\int_{-\infty}^\infty dx \, x
\, n_{\rm loc} (x) =0$. Using Eq.~(\ref{replicas}), ${\mathfrak z}$ can be
expressed in terms of the Fourier transform of $n_{\rm loc}$ as:
\begin{equation} {\mathfrak z} = {\rm e}^{i\frac{2\pi}{L} x_0} \tilde{n}_{\rm
loc} (-\frac{2\pi}{L}) . \label{fourier} \end{equation} If the electron is
localized in a region of space much smaller than $L$, its Fourier transform is
smooth over reciprocal distances of the order of $L^{-1}$ and can be expanded
as: \begin{equation} \tilde{n}_{\rm loc} (-\frac{2\pi}{L}) = 1 - \frac{1}{2}
\left(\frac{2\pi}{L}\right)^2 \int_{-\infty}^\infty dx \, x^2 n_{\rm loc} (x)
+ {\cal O}(L^{-3}) \label{expansion} . \end{equation} A very natural
definition of the center of a localized periodic distribution $|\psi(x)|^2$ is
therefore provided by the phase of ${\mathfrak z}$ as: \begin{equation}
\langle x \rangle = \frac{L}{2\pi} \mbox{Im ln}\, {\mathfrak z} ,
\label{center} \end{equation} which is in fact the formula first proposed by
Selloni {\it et al.}~\cite{Selloni87,brodo}. The expectation value $\langle x
\rangle$ is defined modulo $L$, as expected since $|\psi(x)|^2$ is BvK
periodic.

The above expressions imply $\langle x \rangle \simeq x_0 \; \mbox{mod } L$;
in the special case where $n_{\rm loc}(x)$ can be taken as an even
(centrosymmetric) function, its Fourier transform is real and \equ{fourier}
yields indeed $\langle x \rangle \equiv x_0 \; \mbox{mod } L$. In the case of
extreme delocalization we have instead $|\psi(x)|^2 = 1/L$ and ${\mathfrak z}
= 0$: hence the center of the distribution $\langle x \rangle$, according to
\equ{center}, is ill--defined.  For a more general delocalized state, we expect
that ${\mathfrak z}$ goes to zero at large $L$~\cite{rap107}.

We have therefore arrived at a definition of $\langle x \rangle$ within BvK
which has many of the desirable features we were looking for: nonetheless,
there is a property that is even more important, and which we are going to
demonstrate now. Suppose the potential which the electron moves in has a
slow time dependence---as was the case in Ref.~\cite{Selloni87}---and we wish
to follow the adiabatic evolution of the electronic state $|\psi \rangle$. If
we call $|\varphi_j \rangle$ the instantaneous eigenstates at time $t$, the
lowest order adiabatic evolution of the ground--state density matrix
is~\cite{Thouless}: \[ |\psi \rangle \langle \psi | \simeq |\varphi_0 \rangle
\langle \varphi_0 | + i \sum_{j \neq 0} \left( | \varphi_j \rangle \frac{
\langle \varphi_j | \dot{\varphi}_0 \rangle}{ \epsilon_j - \epsilon_0} \langle
\varphi_0 | - \mbox{Hc} \right), \label{niu} \] where the phases have been
chosen in order to make $|\varphi_0 \rangle$ orthogonal to its time derivative
$ | \dot{\varphi}_0 \rangle$. The macroscopic electrical current flowing
through the system at time $t$ is therefore: \[ \langle j \rangle = -
\frac{1}{L}  \langle \psi | p | \psi \rangle  \simeq - \frac{i}{L} \sum_{j
\neq 0} \frac{ \langle \varphi_0 | p | \varphi_j \rangle \langle \varphi_j |
\dot{\varphi}_0 \rangle}{ \epsilon_j - \epsilon_0} + \mbox{cc} . \] It is then
rather straightforward to prove that $\langle j \rangle$ to lowest order in
$1/L$ equals  $- (1/L) \, d \langle x \rangle / dt$, where $\langle x \rangle$
is evaluated using in \equ{center} the instantaneous ground eigenstate:
\begin{equation} \langle j \rangle \simeq - \frac{1}{2\pi} \mbox{Im }
\frac{d}{dt} \mbox{ln }  \langle \varphi_0 | {\rm e}^{i\frac{2\pi}{L}x} |
\varphi_0 \rangle \label{center2} .  \end{equation} This finding proves the
value of the ``electron--in--broth'' formula, \eqs{z}{center} in studying
electron transport~\cite{Selloni87,brodo}.

\section{The main formula: Many electrons}

So much about the one--electron problem: we are now going to consider a finite
density of electrons in the periodic box. To start with, irrelevant spin
variables will be neglected: for the sake of notation simplicity, I will first
illustrate the main concepts on a system of ``spinless electrons''.  

Even for a system of independent electrons, our approach takes a simple and
compact form if a many--body formulation is adopted. BvK then imposes
periodicity in each electronic variable separately: \begin{equation}
\Psi_0(x_1, \dots, x_i, \dots, x_N) = \Psi_0(x_1, \dots, x_i\! + \! L, \dots,
x_N) .  \label{perio} \end{equation} Our interest is indeed in studying a bulk
system: $N$ electrons in a segment of length $L$, where eventually the
thermodynamic limit is taken: $L \rightarrow \infty$, $N \rightarrow \infty$,
and $N/L = n_0$ constant.  We also assume the ground state nondegenerate, and
we deal with insulating systems only: this means that the gap between the
ground eigenvalue and the excited ones remains finite for $L \rightarrow
\infty$.  

We start defining the one--dimensional analogue of $\hat{\bf R}$, namely, the
multiplicative operator $\hat{X} = \sum_{i=1}^N x_i$, and the complex number
\begin{equation} {\mathfrak z}_N = \langle \Psi | {\rm e}^{i\frac{2\pi}{L}
\hat{X}} | \Psi \rangle .  \label{general} \end{equation} It is obvious that
the operator $\hat{X}$ is ill--defined in our Hilbert space, while its complex
exponential appearing in \equ{general} is well defined. The main result of
Ref.~\cite{rap100} is that the ground--state expectation value of the position
operator is given by the analogue of \equ{center}, namely: \begin{equation}
\langle X \rangle = \frac{L}{2\pi} \mbox{Im ln } {\mathfrak z}_N  ,
\label{main} \end{equation} a quantity defined modulo $L$ as above.  

The right--hand side of Eq.~(\ref{main}) is not simply the expectation value
of an operator: the given form, as the imaginary part of a logarithm, is
indeed essential. Furthermore, its main ingredient is the expectation value
of the multiplicative operator ${\rm e}^{i\frac{2\pi}{L} \hat{X}}$: it is
important to realize that this is a genuine {\it many--body} operator. In
general, one defines an operator to be one--body whenever it is the {\it sum}
of $N$ identical operators, acting on each electronic coordinate separately:
for instance, the $\hat{X}$ operator is such. In order to express the
expectation value of a one--body operator the full many--body wavefunction is
not needed: knowledge of the one--body reduced density matrix $\rho$ is
enough: I stress that, instead, the expectation value of ${\rm
e}^{i\frac{2\pi}{L} \hat{X}}$ over a correlated wavefunction {\it cannot} be
expressed in terms of $\rho$, and knowledge of the $N$-electron wavefunction
is explicitly needed. In the special case of a single--determinant, the
$N$-particle wavefunction is uniquely determined by the one--body reduced
density matrix $\rho$ (which is the projector over the set of the occupied
single--particle orbitals): therefore the expectation value $\langle X
\rangle$, \equ{main}, is uniquely determined by $\rho$. But this is peculiar
to uncorrelated wavefunctions only: this case is discussed in detail below.

As in the one--body case, whenever the many--body Hamiltonian is slowly
varying in time, the macroscopic electrical current flowing through the system
is given by  \[ \langle J \rangle = - \frac1L \frac{d}{dt} \langle X \rangle ,
\label{j} \] where $\langle X \rangle$ is evaluated using in \equ{general} the
instantaneous ground eigenstate of the Hamiltonian at time $t$: this result is
proved in Ref.~\cite{rap100}.  Considering now the limit of a large system,
$\langle X \rangle$ is an extensive quantity: the macroscopic current $\langle
J \rangle$, \equ{j}, goes therefore to a well defined thermodynamic limit.

We stress that nowhere in our presentation have we assumed crystalline
periodicity. Therefore our definition of $\langle X \rangle$ is very general:
it applies to any condensed system, either ordered or disordered, either
independent--electron or correlated.

\section{Macroscopic polarization}

\begin{figure}[b] \begin{center} \setlength{\unitlength}{1cm}
\begin{picture}(14,5.2) \put(0,0){\includegraphics{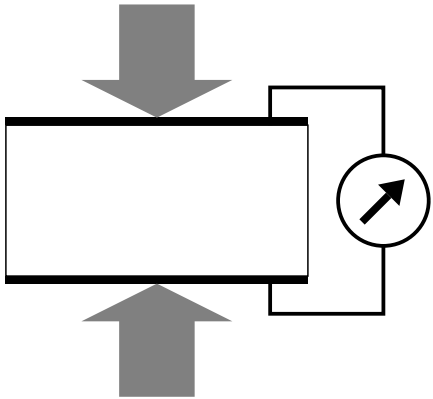}}
\put(7,4){\parbox[t]{7cm}{\caption{\label{fig:piezo} A meaurement of the
piezoelectric effect. The current flowing along a crystal shorted out by a
capacitor is measured while the crystal is strained in one direction.  }}}
\end{picture} \end{center} \end{figure}

In the Introduction, we have discussed what polarization is {\it not}, by
outlining some incorrect definitions~\cite{Kittel,Ashcroft} and their
problems~\cite{Martin74}. We have not stated yet what polarization really {\it
is}: to this aim, a few experimental facts are worth recalling.  The absolute
polarization of a crystal in a given state has never been measured as a bulk
property, independent of sample termination.  Instead, well known bulk
properties are {\it derivatives} of the polarization with respect to suitable
perturbations: permittivity, pyroelectricity, piezoelectricity, dynamical
charges, In one important case---namely, ferroelectricity---the relevant bulk
property is inferred from the measurement of a finite difference (polarization
reversal). In all cases, the derivative or the difference in the polarization
is typically accessed via the measurement of a macroscopic current. For
instance, to measure the piezoelectric effect, the sample is typically
strained along the piezoelectric axis while being shorted out with a capacitor
(see Fig.~\ref{fig:piezo}).  

The theory discussed here  only concerns phenomena where the macroscopic
polarization is induced by a source other than an electric field. Even in this
case, the polarization may (or may not) be {\it accompanied} by a field,
depending on the boundary conditions chosen for the macroscopic sample. The
theory addresses polarization differences {\it in zero field}: this concerns
therefore lattice dynamics, piezoelectricity (as in the ideal experiment
sketched in Fig.~\ref{fig:piezo}), and ferroelectricity. Notably, the theory
reported here {\it does not} address the problem of evaluating the dielectric
constant: this can be done using alternative approaches, such as the
well--established linear--response theory~\cite{rap_a16,Giannozzi91}, or other
more innovative theories~\cite{innovative}.

The bulk quantity of interest, to be compared with experimental measurements,
is the polarization difference between two states of the given solid,
connected by an adiabiatic transformation of the Hamiltonian. The electronic
term in this difference is: \begin{equation} \Delta P = \int_0^{\Delta t}
\!\! dt \; J(t), \label{current} \end{equation} where $J(t)$ is the current
flowing through the sample while the potential is adiabatically varied, {\it
i.e.} precisely the quantity discussed in the previous Section, \equ{j}.
Notice that in the adiabatic limit $\Delta t$ goes to infinity and $J(t)$
goes to zero, while \equ{current} yields a finite value, which only depends
on the initial and final states.  We may therefore write: \begin{equation} P
= - \lim_{L \rightarrow \infty} \frac{1}{2\pi} \mbox{Im ln }  \langle \Psi |
{\rm e}^{i\frac{2\pi}{L} \hat{X}} | \Psi \rangle , \label{limit}
\end{equation} where it is understood that \equ{limit} is to be used twice,
with the final and with the initial ground states, in order to evaluate the
quantity of interest $\Delta P$. Notice that $L \rightarrow \infty$ in
\equ{limit} is a rather unconventional limit, since the exponential operator
goes formally to the identity, but the size of the system and the number of
electrons in the wavefunction increase with $L$.

\section{The case of independent electrons}

We now specialize to an uncorrelated system of independent electrons, whose
$N$-electron wavefunction $| \Psi \rangle$ is a Slater determinant.  As
discussed above, in this case the expectation value $\langle X \rangle$,
Eq.~(\ref{main}), is uniquely determined by the one--body density matrix.
However, the formulation is simpler when expressing $\langle X \rangle$ and
the resulting polarization $P$ directly in terms of the orbitals.

We restore explicit spin variables from now on. Suppose $N$ is even, and $|
\Psi \rangle$ is a singlet.  The Slater determinant has thus the form:
\begin{equation} | \Psi \rangle = \frac{1}{\sqrt{N!}} |\varphi_1
\overline{\varphi}_1 \varphi_2 \overline{\varphi}_2 \dots \varphi_{N/2}
\overline{\varphi}_{N/2}| , \label{slater} \end{equation} where $\varphi_i$
are the single-particle orbitals. It is then expedient to define
\begin{equation} | \tilde{\Psi} \rangle =  {\rm e}^{i\frac{2\pi}{L} \hat{X}} |
\Psi \rangle : \label{tilde} \end{equation} even $| \tilde{\Psi} \rangle$ is
indeed a Slater determinant, where each orbital $\varphi_i(x)$ of $| \Psi
\rangle$ is multiplied by the plane wave $ {\rm e}^{i\frac{2\pi}{L} x}$. 
According to a well known theorem, the overlap amongst two determinants is
equal to the determinant of the overlap matrix amongst the orbitals. We
therefore define the matrix (of size $N/2\times N/2$): \begin{equation} S_{ij}
= \langle \varphi_i | {\rm e}^{i\frac{2\pi}{L} x} | \varphi_j \rangle =
\int_0^L \! dx \, \varphi^*_i(x) {\rm e}^{i\frac{2\pi}{L} x} \varphi_j(x),
\label{overlap} \end{equation} in terms of which we easily get
\begin{equation} P = - \frac{1}{2\pi} \mbox{Im ln } \langle \Psi |
\tilde{\Psi} \rangle = - \frac{1}{\pi} \mbox{Im ln det } S , \label{single}
\end{equation} where the factor of 2 accounts for double spin occupancy, and
the expression becomes accurate in the limit of a large system.

The expression of \equ{single} goes under the name  of ``single--point Berry
phase'' (almost an oxymoron!), and was first proposed by the present author in
a volume of lecture notes~\cite{rap_a17}.  Since then, its three-dimensional
generalization has been used in a series of DFT calculations for
noncrystalline systems~\cite{simulations}, and  has been scrutinized in some
detail in Ref.~\cite{rap101}.

The case of a crystalline system of independent electrons is the one which
historically has been solved first~\cite{rap73,rap78,King93}, though along a
very different logical path~\cite{rap_a12,rap_a16,Martin97a,rap_a18} than
adopted here. I am going to outline how the present formalism leads to the
earlier results.

For the sake of simplifying notations, I am going to consider the case of an
insulator having only one completely occupied band. The single--particle
orbitals may be chosen in the Bloch form: indeed, the canonical ones {\it
must} have the Bloch form. The many--body wavefunction, \equ{slater}, becomes
in the crystalline case: \[ | \Psi \rangle = \frac{1}{\sqrt{N!}} |\psi_{q_1
}\overline{\psi}_{q_1} \psi_{q_2} \overline{\psi}_{q_2} \dots \psi_{q_{N/2}}
\overline{\psi}_{q_{N/2}}| .  \label{slater2} \end{equation} The lattice
constant is $a = 2L/N$, and the Bloch vectors entering \equ{slater2} are
equally spaced in the reciprocal cell $(0,  2 \pi / a]$: \begin{equation} q_s
= \frac{4 \pi}{N a} s, \quad s=1,2, \dots, N/2 .  \label{points} \end{equation}

Owing to the orthogonality properties of the Bloch functions, the overlap
matrix elements in \equ{overlap} vanish except when $q_{s'} = q_s - 2\pi /L$,
that is $s' = s\!-\!1$: therefore the determinant in \equ{single} factors as
\[ P  = \frac{1}{\pi} \mbox{Im ln } \prod_{s=1}^{N/2} \langle \psi_{q_s} | {\rm
e}^{i \frac{2\pi}{L} x} | \psi_{q_{s-1}} \rangle , \label{berry} \] where 
$\psi_{q_{0}}(x) \equiv \psi_{q_{N/2}}(x)$ is implicitly understood
(so--called periodic gauge).

The expression in \equ{berry} is precisely the one first proposed by
King--Smith and Vanderbilt~\cite{King93} as a discretized form for the Berry
phase: in fact, the multi--band three--dimensional generalization of
\equ{berry} is the standard formula~\cite{rap_a12,rap_a16} implemented in
first--principle calculations~\cite{William} of macroscopic polarization in
crystalline dielectrics, either within DFT or HF.

\section{Critical rethinking of DFT}

The modern viewpoint about macroscopic polarization has even spawned a
critical rethinking of density--functional theory in extended systems. The
debate started in 1995 with a paper by Gonze, Ghosez, and
Godby~\cite{Gonze95}, and continues these days~\cite{debate}. The treatment of
polarization provided in the present work allows discussing the issue in a
very simple way.

The celebrated Hohenberg--Kohn (HK) theorem, upon which DFT is
founded~\cite{DFT}, states that there exists a universal functional $F[n]$,
which determines the exact ground--state energy and all other ground--state
properties of the system.  The main hypotheses are that the magnetic field is
vanishing, the ground state is non degenerate, and---most important to the
present purposes---the system is {\it finite}, with a square--integrable
ground wavefunction. This is precisely the key point when dealing with an
extended system.  

Ideally, it is possible to refer to a macroscopic but finite system:
polarization is then by definition the dipole divided by the volume, and
\equ{dipole} safely applies. As a consequence, the polarization of the real
interacting system is identical to the one of the fictitious noninteracting
Kohn--Sham (KS) system. Unfortunately, such polarization depends on the charge
distribution both in the bulk and at the surface: according to
Ref.~\cite{Gonze95}, this fact implies a possible ``ultranonlocality'' in the
KS potential: two systems having the same density in the bulk region may have
qualitatively different KS potentials in the same region, and different
polarizations as well. The issue can be formulated in a more transparent way
by recasting it in the language of the present work.

As discussed in the Introduction, condensed matter theory invariably works in
a different way: one adopts BvK since the very beginning, and the system has
no surface by construction. The original HK theorem was formulated for the
case where the Schr\"odinger's equation is solved imposing square--integrable
boundary conditions, but the same theorem holds within BvK, with an identical
proof, for a {\it finite} $N$-electron system. We can then define, even within
BvK, the fictitious KS system of noninteracting electrons, having the same
density as the interacting one: if the system is crystalline, the KS orbitals
have the Bloch form.

Now the question becomes: do the interacting system and the corresponding KS
noninteracting one have the same polarization? The answer is actually ``no'':
in fact, within BvK, polarization is {\it not} a function of the density, not
even of the one--body density matrix, as stressed above.  Numerical evidence
of the fact that the two polarizations are not equal has been
given~\cite{debate}.  

Other important implications concern the occurrence of macroscopic electric
fields within DFT: we refer to the original literature~\cite{Gonze95,debate}
about this issue, while here we limit ourselves to just remarking an important
point. The potential (both one--body and two--body) within the Schr\"odinger's
equation must be BvK periodic, otherwise the Hamiltonian is an ill--defined
operator in the Hilbert space. The periodicity of the potential is tantamount
to enforcing a vanishing macroscopic electric field: therefore some {\it
ad--hoc} strategies must be devised in order to cope with nonzero electric
fields, as is indeed done in Refs.~\cite{Gonze95,debate}.

\section*{Acknowledgments}

Invaluable discussions with R.M. Martin, G. Ort\'{\i}z, and D. Vanderbilt are
gratefully acknowledged. Part of this work has been performed while attending
the 1998 workshop ``Physics of Insulators'' at the Aspen Center for Physics.
Partially supported by ONR grant N00014-96-1-0689.

\end {document}